\documentclass[fleqn,twoside]{article}
\usepackage{espcrc2}

\usepackage{graphicx}
\usepackage[figuresright]{rotating}

\usepackage[dvips]{color}

\newcommand{\AmS}{{\protect\the\textfont2
  A\kern-.1667em\lower.5ex\hbox{M}\kern-.125emS}}

\newcommand{\Journal}[4]{{ #1} {\bf #2} {(#3)} {#4}}

\newcommand{\PL}{Phys. Lett.}
\newcommand{\PR}{Phys. Rev.}

\newcommand{\PRL}{Phys. Rev. Lett.}

\title{Lepton Flavor Violation in a Long Baseline Experiment}

\author{Toshihiko Ota\address{Department of Physics, Osaka University,
        Machikaneyama 1-1, Toyonaka, Osaka, 560-0043, Japan}
       and
       Joe Sato\address{Department of Physics, Saitama University,
       Shimo-Okubo 255, Sakura-ku, Saitama, 338-8570, Japan}
       \address{
       Institut f\"ur Theoretische Physik, Physik-Department, 
       Technische Universitaet M\"unchen, 
       James-Franck-Strasse D-85748 Garching, Germany}
       }
       
\begin{document}

\begin{abstract}
We evaluate the size of a coupling for a lepton flavor violating interaction 
with neutrinos (nLFV) in the minimal supersymmetric standard model 
with right-handed neutrinos (MSSMRN).
It is expected that the nLFV interactions are detected  
in a long baseline neutrino oscillation experiment.
We here deal with 
$\mu^{-} \rightarrow e^{-} \bar{\nu}_{e} \nu_{\tau}$
in $\nu_{\mu} \rightarrow \nu_{\tau}$ channel at a neutrino factory. 
First, we show the numerical study on feasibility to observe such 
interaction   
and point out that we can search for them with some oscillation channels.
Next, we make a numerical calculation for the size of the nLFV
coupling in the framework of the MSSMRN. 
A long baseline experiment is sensitive not only to the lepton mixings
in the standard model but also to new physics effect. 
\vspace{1pc}
\end{abstract}

\maketitle

\section{Introduction}

The processes of the lepton flavor violation in the charged lepton sector (cLFV)
are the clear signature beyond the standard model (SM), 
and they are searched for by many experiments.
In the minimal supersymmetric standard model with right-handed neutrinos
(MSSMRN),
the effective cLFV interactions are originated from
the off-diagonal elements of the left-handed slepton mass 
matrix, $(m_{\tilde{L}}^{2})_{ij}$ $(i\neq j)$, 
which are induced by the radiative corrections with neutrino
Yukawa interaction
\cite{BM,HMTYY,JoeTobe}. 
The size of the cLFV processes in many classes of such a model 
can reach the sensitivity of the forthcoming experiments.
Therefore, the cLFV search is one of the promising experiments to
find the new physics evidence.
However, we here propose an alternative method to do it, 
the search for the processes of the lepton flavor violation with
neutrinos (nLFV) in a long baseline neutrino oscillation experiment.
The feasibility study on the nLFV interaction search without the specific
models is made 
by Refs.\cite{Grossman,GGGN,NewPhysMatter,HV,HSV,OSY,OS}.
It is pointed out that 
the nLFV signal is enhanced by 
the interference between the amplitude
with the nLFV interactions and that of 
the standard oscillation \cite{HSV,OSY,OS}.

We here investigate the nLFV interaction in the MSSMRN.
Since the origin of the nLFV processes is the same as that of the cLFV
processes and, in addition, there is an enhancement mechanism due to 
the interference, the detectable nLFV coupling can be expected.
We show the numerical result on the size of the nLFV coupling and the
correlation with the cLFV process.

In the next section, the enhancement mechanism is explained and 
the way to parameterize the nLFV effect in an oscillation experiment
is shown.
We here deal with the nLFV process, $\mu^{-} \rightarrow e^{-}
\bar{\nu}_{e} \nu_{\tau}$ in $\nu_{\mu} \rightarrow \nu_{\tau}$ channel
at a neutrino factory.
In Sec.3, we evaluate numerically the size of the effective coupling 
for the nLFV interaction in the MSSMRN. 
Finally, the conclusions are given in Sec.4.

\section{Enhancement mechanism in nLFV interaction search}
In the neutrino oscillation experiments,
we only observe the charged particles which interact with the neutrino,
namely, the neutrino appears only in the intermediate state.
Therefore, 
the existence of the nLFV interactions suggests that there are some
amplitudes whose initial and final states are the same as the standard
oscillation amplitude.
Let us show one example.
In $\nu_{\mu} \rightarrow \nu_{\tau}$ oscillation measurement at a neutrino
factory,
we can only know the decay of the muon at the muon storage ring 
and the appearance of the tau lepton at the detector.
These observations are interpreted by the standard oscillation scenario as
the muon decays into a muon neutrino and it oscillates to the tau neutrino
and it produces a tau lepton through the charged current interaction:  
\begin{eqnarray}
&&A_{\rm SM}(\mu^{-} + I \rightarrow \tau^{-} + F)  \\
&=&\hspace{-0.3cm}
A(\mu^{-} \rightarrow \nu_{\mu} \bar{\nu}_{e} e^{-}) 
A(\nu_{\mu} \rightarrow \nu_{\tau})
A(\nu_{\tau}  d \rightarrow \tau^{-} u). \nonumber
\end{eqnarray}
However, the same result can be reproduced by the other process
which includes the nLFV interaction, 
the muon decays into a tau neutrino:
\begin{eqnarray}
&&A_{\rm nLFV}(\mu^{-} + I \rightarrow \tau^{-} + F)  \\
&=&\hspace{-0.3cm}
A(\mu^{-} \rightarrow \nu_{\tau} \bar{\nu}_{e} e^{-}) 
A(\nu_{\tau} \rightarrow \nu_{\tau})
A(\nu_{\tau}  d \rightarrow \tau^{-} u). \nonumber
\end{eqnarray}
The oscillation probability for $\nu_{\mu} \rightarrow \nu_{\tau}$
is modified to
\begin{eqnarray}
P_{\mu^{-} \rightarrow \tau^{-}}
= |A_{\rm SM}|^{2} +
2 {\rm Re}[A_{\rm SM}^{*} A_{\rm nLFV}] + \cdots.
\end{eqnarray}
The second term comes from the interference effect, which is proportional to the
nLFV coupling itself not its square.
We define $\epsilon^{s}_{\mu\tau}$ as the ratio between the coupling of 
the standard interaction
and the nLFV interaction\footnote{%
Here, we only consider the $(V-A)(V-A)$ type nLFV interaction.}:
\begin{equation}
\epsilon^{s}_{\mu \tau} \equiv
A(\mu^{-} \rightarrow \nu_{\tau} \bar{\nu}_{e} e^{-})
\bigl/
A(\mu^{-} \rightarrow \nu_{\mu} \bar{\nu}_{e} e^{-}).
\end{equation}
We can deal with this effect by treating 
the initial neutrino state (the source state) as the flavor mixture state
\cite{Grossman}:
\begin{equation}
|\nu_{\mu}^{s} \rangle = |\nu_{\mu} \rangle 
+ \epsilon^{s}_{\mu\tau} |\nu_{\tau} \rangle.
\end{equation}
The other types of nLFV interaction can enter the neutrino propagation process
as the exotic matter effect potential
\cite{GGGN,NewPhysMatter,HV,HSV,OSY,OS}
and also the neutrino detection process \cite{Grossman}.
They are investigated in our future work,
we here concentrate our attention on the nLFV interaction at the source state.

\begin{figure}[htb]
\unitlength=1cm
\begin{picture}(7,6)
\hspace{0.5cm}
\includegraphics[width=6cm]{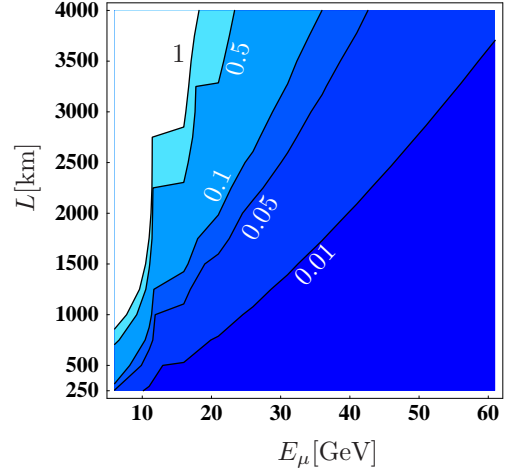}
 \put(-6.5,3){\rotatebox{90}{$L$[km]}}
 \put(-3,-0.3){$E_{\mu}$[GeV]}
 \put(-4.4,5){1}
 \color[rgb]{1,1,1}
 \put(-3.7,4.8){\rotatebox{75}{0.5}}
 \put(-4,3.1){\rotatebox{65}{0.1}}
 \put(-3.5,2.6){\rotatebox{58}{0.05}}
 \put(-2.8,2){\rotatebox{47}{0.01}}
\end{picture}\vspace{-0.5cm}
 \caption{The necessary exposure (= muon number $\times$ detector size)
 to detect the nLFV interaction  
 $\epsilon^{s}_{\mu\tau} = 3.0 \times 10^{-3} {\rm e}^{{\rm i} (\pi/2)}$
 at 90\% CL in $\nu_{\mu} \rightarrow \nu_{\tau}$ channel \cite{OSY}.
 The numbers in the figure are normalized by $10^{21}$ muons $\times$ 100
 kt detector.}
\label{Fig:feasibility}
\end{figure}

Figure \ref{Fig:feasibility} shows the sensitivity to the nLFV
interaction $\epsilon^{s}_{\mu\tau}$.
In this numerical study, we take 10\% uncertainty for the mixing angles
and the mass square differences and we treat the CP phase in the
Maki-Nakagawa-Sakata (MNS) matrix as a freely fitting parameter.
The nLFV signal is robust against the adjustment of the standard
oscillation parameters because of its characteristic energy dependence
\cite{OSY}.
This plot indicates that 
the nLFV interaction with $\epsilon^{s}_{\mu\tau} =
\mathcal{O}(10^{-4})$
can be observed at a neutrino factory in this oscillation channel.

\section{Size of the nLFV coupling in MSSMRN}
\begin{figure}[htb]
\hspace{0.8cm}
\unitlength=0.6cm
\begin{picture}(10,5.5)
\thicklines
\put(0.5,4){\line(1,0){2}}
\put(1.5,4){\vector(1,0){0}}
\put(0.5,4.3){$\mu^{-}$}
\put(2.5,4){\circle*{0.2}}
\multiput(2.5,4)(0.5,0){8}{\line(1,0){0.3}}
\put(3.5,4){\vector(1,0){0}}
\put(3.3,4.3){\footnotesize $\tilde{l}^{-}$}
\put(4.5,4){\circle*{0.2}}
\put(5.5,4){\vector(1,0){0}}
\put(5.2,4.3){\footnotesize $\tilde{\nu}$}
\put(6.5,4){\circle*{0.2}}
\put(6.5,4){\line(1,0){2}}
\put(8,4){\vector(1,0){0}}
\put(8,4.3){$\nu_{\tau} $}
\put(4.5,4){
\qbezier(-2,0)(-2.01,0.35)(-1.88,0.68)
\qbezier(-1.88,0.68)(-1.78,1.03)(-1.53,1.29)
\qbezier(-1.53,1.29)(-1.3,1.55)(-1,1.73)
\qbezier(-1,1.73)(-0.69,1.9)(-0.35,1.97)
\qbezier(-0.35,1.97)(0,2.03)(0.35,1.97)
\qbezier(2,0)(2.01,0.35)(1.88,0.68)
\qbezier(1.88,0.68)(1.78,1.03)(1.53,1.29)
\qbezier(1.53,1.29)(1.3,1.55)(1,1.73)
\qbezier(1,1.73)(0.69,1.9)(0.35,1.97)
}
\put(4.8,6){\vector(1,0){0}}
\put(4.5,4){
\qbezier(-2,0)(-2.36,0.417)(-1.88,0.68)
\qbezier(-1.88,0.68)(-1.47,0.85)(-1.53,1.29)
\qbezier(-1.53,1.29)(-1.54,1.84)(-1,1.73)
\qbezier(-1,1.73)(-0.58,1.6)(-0.35,1.97)
\qbezier(-0.35,1.97)(0,2.35)(0.35,1.97)
\qbezier(2,0)(2.36,0.417)(1.88,0.68)
\qbezier(1.88,0.68)(1.47,0.85)(1.53,1.29)
\qbezier(1.53,1.29)(1.54,1.84)(1,1.73)
\qbezier(1,1.73)(0.58,1.6)(0.35,1.97)
}
\put(4.3,6.5){$\tilde{\chi}^{0}$}
\rotatebox{30}{
\multiput(5.2,1.5)(0,-1){2}
{\qbezier(0,0)(-0.25,-0.25)(0,-0.5)
 \qbezier(0,-0.5)(0.25,-0.75)(0,-1.0)}
}
\put(4.5,2.5){$W^{-}$}
\end{picture}\vspace{-1.8cm}
\caption{One example of the diagrams which induce the effective
 interaction $\mu^{-} \rightarrow \nu_{\tau} \bar{\nu}_{e} e^{-} $.}
\label{Fig:diagram}
\end{figure}
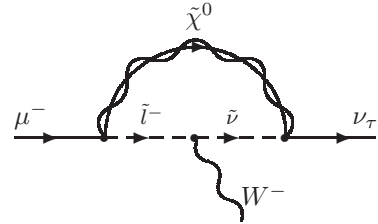
In this section, we turn to the theoretical prediction on the size of
$\epsilon^{s}_{\mu\tau}$ in the MSSMRN.
In this model, the nLFV coupling $\epsilon^{s}_{\mu\tau}$ 
is induced by the radiative correction.
One example is shown in Fig.\ref{Fig:diagram}.
We can roughly estimate the order of $\epsilon^{s}_{\mu\tau}$
as
\begin{equation}
\epsilon^{s}_{\mu\tau}
\sim
\frac{g^{2}}{(4\pi)^{2}}
\frac{(m_{\tilde{L}}^{2})_{32}}{m_{\rm SUSY}^{2}},
\end{equation}
where $m_{\rm SUSY}^{}$ is the typical soft SUSY breaking scale.
It is well known that the branching fraction of the cLFV process 
$\tau\rightarrow \mu\gamma$ can be approximated  as \cite{HMTYY}
\begin{equation}
{\rm Br}(\tau\rightarrow\mu\gamma)
\propto
\frac{\alpha^{3}}{G_{F}^{2}} \frac{|(m_{\tilde{L}}^{2})_{32}|^{2}}{m_{\rm SUSY}^{8}}.
\end{equation} 
These equations lead to the correlation between the cLFV process and the
nLFV process: 
\begin{equation}
\epsilon^{s}_{\mu\tau}
 \propto 
 \sqrt{{\rm Br}(\tau\rightarrow\mu\gamma)} \times m_{\rm SUSY}^{2}.
 \label{eq:correlation}
\end{equation}
\begin{figure}[htb]
\hspace{0.5cm}
\unitlength=1cm
\begin{picture}(7,5.5)
 \includegraphics[width=6cm]{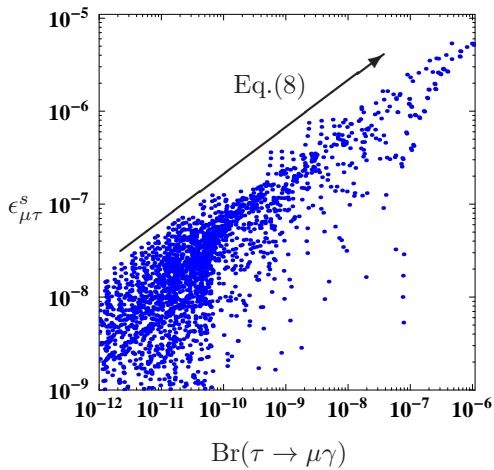}
 \put(-6.5,3){$\epsilon^{s}_{\mu\tau}$}
 \put(-3.8,-0.3){${\rm Br}(\tau\rightarrow \mu\gamma)$}
 \thicklines
 \put(-5,2.5){\vector(4,3){3.5}}
 \put(-3.5,4.6){Eq.(\ref{eq:correlation})}
\end{picture}
\caption{Correlation between $\epsilon^{s}_{\mu\tau}$ and 
 ${\rm Br(\tau\rightarrow \mu\gamma)}$.
 In this calculation, we scan the universal mass parameter with some
 different texture of the neutrino Yukawa matrix.
 We fix $\tan\beta=10$ and $\mu>0$.}
\label{Fig:scatter-plot}
\end{figure}
The numerical result is shown in Fig.\ref{Fig:scatter-plot}.
In this calculation, we scan so called the universal soft mass
parameters and the texture of the neutrino Yukawa matrix.
The detail is shown in Ref.\cite{OS-Future}.  
This result can be easily understood by
Eq.(\ref{eq:correlation}). 

\section{Conclusions}
We investigate the size of the nLFV interaction $\mu^{-} \rightarrow
\nu_{\tau} \bar{\nu}_{e} e^{-}$ in the framework of the
MSSMRN.
In this model, the cLFV process such as $\tau\rightarrow\mu\gamma$ can 
be large enough to be detected by the forthcoming experiments.
Since the origin of the cLFV interactions and that of nLFV interactions
are the same, and 
in addition, there is an enhancement mechanism 
for the nLFV signal due to the interference effect, 
the sizable nLFV effect can be expected.
The sensitivity to $\epsilon^{s}_{\mu\tau}$ was evaluated as 
$\mathcal{O}(10^{-4})$ in Ref.\cite{OSY}.
We here 
evaluate the size of the nLFV coupling $\epsilon^{s}_{\mu\tau}$
in the MSSMRN and show the correlation with the cLFV process 
$\tau \rightarrow \mu \gamma$.
We may have a chance not only to measure the oscillation parameters in
the SM but also to detect the new physics effect at a neutrino factory.



\end{document}